\documentclass[11pt,a4paper]{article}

\newlength{\dinwidth}
\newlength{\dinmargin}
\usepackage{epsfig}
\usepackage{graphicx}
\usepackage{cite}
\usepackage{amsmath}
\usepackage{amssymb}
\usepackage{lscape}

\newcommand{\comment}[1]  {}
\newcommand\ie{{\textsl{i.e.\,}}}
\newcommand\eg{{\textsl{e.g.\,}}}
\def\BE{\begin{equation}}
\def\EE{\end{equation}}
\def\BEA{\begin{eqnarray}}
\def\EEA{\end{eqnarray}}

\newcommand\vn{{\bf n}}

\newcommand\vr{{\bf r}}
\newcommand\vs{{\bf s}}
\newcommand\vt{{\bf t}}

\newcommand\vx{{\bf x}}

\newcommand\vz{{\bf z}}
\newcommand\mA{{\bf A}} 
\newcommand\mB{{\bf B}}

\newcommand\mH{{\bf H}}

\newcommand\mM{{\bf M}}

\def\thefootnote{\fnsymbol{footnote}}

\begin{document}

\title{\textbf{Parallel vs. Sequential Belief Propagation Decoding of LDPC Codes over $GF(q)$\\and Markov Sources}}


\author{Nadav Yacov\footnote{Minerva Center and Department of Physics, Bar-Ilan University, Ramat-Gan 52900, Israel
        (e-mail: \{nadav.yacov,hadaref,kanter\}@mail.biu.ac.il).}\,, Hadar Efraim$^{\ast}$, Haggai Kfir\footnote{Minerva Center and Department of
Physics, Bar-Ilan University. Currently with the Israel Aircraft
Industries (IAI) (e-mail: hkfir@iai.co.il).}\,, Ido
Kanter$^{\ast}$ and Ori Shental\footnote{Department of Electrical
Engineering-Systems, Tel-Aviv University, Tel Aviv 69978, Israel
(e-mail:
shentalo@eng.tau.ac.il).}\,\,\,\footnote{Corresponding author\newline Submitted to IEEE Communications Letters, \today. }
}

\date{}
\maketitle

\begin{abstract}
A sequential updating scheme (SUS) for belief propagation (BP)
decoding of LDPC codes over Galois fields, $GF(q)$, and correlated
Markov sources is proposed, and compared with the standard
parallel updating scheme (PUS). A thorough experimental study of
various transmission settings indicates that the convergence rate,
in iterations, of the BP algorithm (and subsequently its
complexity) for the SUS is about one half of that for the PUS,
independent of the finite field size $q$. Moreover, this $1/2$
factor appears regardless of the correlations of the source and
the channel's noise model, while the error correction performance
remains unchanged. These results may imply on the 'universality'
of the one half convergence speed-up of SUS decoding.
\end{abstract}

\renewcommand{\thefootnote}{\arabic{footnote}}
\setcounter{footnote}{0}

\newpage
\section{Introduction}
\comment{\PARstart{L}{ow density}} Low density parity check (LDPC)
codes, first invented by Gallager in 1962~\cite{BibDB:Gallager}
and long afterwards rediscovered in the seminal work of MacKay and
Neal (MN,~\cite{BibDB:MackayNeal}), play a fundamental role in
modern communications, primarily due to their near-Shannon limit
performance. An almost optimal, yet tractable
decoding~\cite{BibDB:MackayNeal} of this class of codes is
empowered by the renowned probabilistic message passing algorithm
of belief propagation (BP,~\cite{BibDB:BookPearl}).

It was also shown~\cite{BibDB:DaveyMackay} that the remarkable
error performance of Gallager's binary LDPC codes can be
significantly enhanced even further by a generalization to higher
finite Galois fields, $GF(q)$ with $\mathbb{Z}\ni q>2$. This
behavior can be rationalized by the fact that the graphical
representation of such codes has less edges and subsequently
relatively longer loops.

As BP is an iterative algorithm, its convergence rate is also a
crucial benchmark in its implementation as a decoder. In
principle, as the communication channel becomes noisier the
decoding time increases, since the latter is dictated by the total
number of BP message passing iterations required for
convergence~\cite{BibDB:RichardsonEtAl}. Furthermore, this number
of iterations tends to diverge when approaching the channel's
capacity~\cite{BibDB:KanterSaad}.

Kfir and Kanter~\cite{BibDB:KfirKanter} were the first to
introduce a serialization method which was shown, by providing
convincing empirical results, to yield half the decoding time and
complexity w.r.t. parallel (flooding) scheduling, while the error
performance does not deteriorate. Following, several sequential
(serial) BP message passing schedules were recently introduced
(\eg,~\cite{BibDB:ReportZhangEtAl} and references therein). It was
shown, either via semi-analytical
methods~\cite{BibDB:ReportZhangEtAl,BibDB:TongWang,BibDB:SharonEtAl}
or by simulations~\cite{BibDB:ZhangFossorier}, that such
sequential schedules converge faster than the standard parallel
schedule.

Despite the aforementioned contributions addressing binary LDPC
codes, there has been no examination of the effect of
serialization on LDPC codes over $GF(q)$.
In this letter, based on a thorough experimental study of a
$GF(q)$ extension of the serialization scheme originally proposed
by Kfir and Kanter~\cite{BibDB:KfirKanter}, we find that not only
sequential decoding over $GF(q)$ accelerates BP convergence w.r.t.
standard flooding, but interestingly the same $1/2$ convergence
ratio arises. The error correction performance is roughly
preserved.

In addition, the convergence of sequential decoding is
investigated for correlated information sources, an issue yet to
be discussed in the literature, addressing only independent and
identically distributed (i.i.d.) sources. Here, the source is
modelled by a Markov process, while a method of dynamical block
priors\cite{BibDB:KfirEtAl} is incorporated within the BP decoding
in order to exploit the prior knowledge on the source statistics
and form a joint source-channel decoding scheme (the move to
$GF(q)$ enables the treatment of Markov sequences with a richer
alphabet). Again, the same factor $1/2$ emerges in this case too,
accelerating the convergence of sequential BP decoding
substantially. These results are corroborated via simulations for
the binary symmetric channel (BSC), binary erasure channel (BEC)
and the binary-input additive white Gaussian noise (BI-AWGN)
channel.

To sum up, the one half factor is found to be robust to the
following extensions: 1) extension of binary sources, $q=2$, to
higher finite field, $q>2$, 2) extension of BP for i.i.d. sources
to the case of BP with dynamical priors used for joint
source-channel decoding, and 3) extension of the BSC case to other
popular channel models. This extension may imply on the
'universality' of the $1/2$ convergence speed-up ratio of
sequential BP decoding.

\section{Sequential and Parallel Joint Source-Channel Decoding}
Consider a MN code with two sparse matrices known both to the
transmitter and the receiver, $\mA (M\times N)$ and $\mB (M\times
M)$, where the indices $N,M$ are the source block length and the
transmitted block length, respectively. All non-zero elements in
$\mA$ and $\mB$ are taken randomly from $\{1,2,\ldots,q-1\}\in
GF(q)$, and $B$ must be invertible.

A proper construction of these matrices is crucial in order to
ensure capacity-achieving performance. In this work we follow the
Kanter-Saad (KS,~\cite{BibDB:KanterSaad}) construction, which
yields very sparse, simple to construct matrices, known to perform
very close to the bound.

Thus encoding a source vector $\vs$ into a codeword $\vt$, with
rate $R=N/M$, is performed by \BE\label{maket}
    \vt=\mB^{-1}\mA\vs \mod q,
\EE where $\vt$ is converted to binary representation and
transmitted over the channel. During transmission, the coded
information $\vt$ is corrupted by a noise vector $\vn$, resulting
in the received vector $\vr=\vt+\vn \mod 2$.\footnote{Hereinafter,
for exposition purposes, a BSC with flip rate $f$ is assumed,
although an extension to other channel models is straightforward
and results for such are addressed in the following.}

Upon receipt, the decoder reconverts $\vr$ back to the original
field and computes the syndrome $\vz=\mB\vr$, which can be
reformulated as \BE
    \vz=\mB(\mB^{-1}\mA
    \vs+\vn)=[\mA\mB]\vx\triangleq\mH\vx \mod q, \label{z and stuff} \EE
where the operator $[\cdot]$ denotes appending of matrices, and
vector $\vx$ is a concatenation of $\vs$ and $\vn$. The decoding
problem is solved efficiently using the BP algorithm, as follows.

The non-zero elements in a row $i$ of the matrix $\mH$ represent
the bits of vector $\vx$ participating in the corresponding check,
$z_{i}$. The non-zero elements in column $j$ represent the checks
in which the $j$'th bit participates. For each non-zero element in
\mH, the algorithm calculates different types of coefficients.

The coefficient $q^{a}_{ij}$ stands for the probability that the
bit $x_{j}$ is $a\in GF(q)$, taking into account the information
of all checks in which it participates, except for the $i$'th
check. The coefficient $r_{ij}^{a}$ indicates the probability that
the bit $x_{j}$ is $a$, taking into account the information of all
bits participating in the $i$'th check, except for the $j$'th bit.

The parallel updating scheme (PUS) consists of alternating
horizontal and vertical passes over the $\mH$ matrix. Each pair of
horizontal and vertical passes is defined as an iteration. In the
horizontal pass, all the $r_{ij}^{a}$ coefficients are updated,
row after row, by \BE\label{eq_r} r^a_{ij}=\sum _{(all\,
configurations\, with\, x_{j}=a,\, satisfing\, z_{i})}\prod
_{j'\neq j}q_{ij'}^{x_{j'}}, \EE where it is clear that the
multiplication is performed only over the non-zero elements of the
matrix $\mH$.

In the vertical pass, all $ q^{a}_{ij}$ are computed, column by
column, using the updated values of $ r^{a}_{ij}$ \BE \label{eq_q}
q_{ij}^{a}=\alpha_{ij}p_{j}^{a}\prod_{i'\neq i}r_{i'j}^{a}, \EE
where $\alpha _{ij}$ is a normalization factor such that
$\sum_{a=1}^{q}q_{ij}^{a}=1$, and $p_{j}^{a}$ represents the prior
knowledge about bit $j$ being in state $a$. Now the
pseudo-posterior probability can be computed by \BE \label{eq_Q}
Q_{j}^{a}=\alpha_{j}p_{j}^{a}\prod _{i}r_{ij}^{a}.\EE Again,
$\alpha _{j}$ is a normalization constant satisfying
$\sum_{a=1}^{q}Q^{a}_{j}=1 $, and $i$ runs only over non-zero
elements of $\mH$. Each iteration ends by generating the estimates
$\hat{\vx}$ by clipping the $Q_{j}$'s.

At the end of each iteration a convergence test, checking if
$\hat{\vx}$ solves $\mH\hat{\vx}=\vz$, is performed. If some of
the $M$ equations are violated, the algorithm turns to the next
iteration until a pre-defined maximal number of iterations is
reached with no convergence. Note that there is no inter-iteration
information exchange between the bits: all $r_{ij}^{a}$ values are
updated using the previous iteration data.

In the proposed sequential updating scheme (SUS), we perform the
horizontal and vertical passes separately for each bit in $\vx$. A
single sequential iteration for the bit $x_{j}$ consists of the
following steps:

\begin{enumerate}
\item For a given $j$ all $r_{ij}^{a}$ are updated. More
precisely, for all non-zero elements in column $j$ of $\mH$, use
Eq.~(\ref{eq_r}) for updating $r_{ij}$. Note that this is only a
\emph{partial} horizontal pass, since only $r_{ij}^{a}$'s
belonging to a specific column are updated. \item After all $
r_{ij}^{a}$'s belonging to a column $j$ are updated, a vertical
pass as defined in Eq.~(\ref{eq_q}) is performed over column $j$.
Again, this is a \emph{partial} vertical pass, referring only to
one column. \item Steps 1-2 are repeated for the next column,
until all columns in $\mH$ are updated. \item Finally, the
pseudo-posterior probability value $Q_{j}$, is computed by
Eq.~(\ref{eq_Q}).
\end{enumerate}
After all variable nodes are updated, the algorithm continues as
for the parallel scheme: clipping, checking the validity of the
$M$ equations and proceeding to the next iteration.

As for the priors, side information on the Markovian nature of the
source is dynamically incorporated within the decoding process.
During the vertical pass, a prior knowledge is assigned to each
decoded symbol according to the assumed statistics (evidently, for
the i.i.d. case this would simply be $\Pr(s=a)=1/q$ for all the
source symbols.)

The key here is that one can re-estimate and re-assign these
priors after every iteration. Consider, for instance, three
successive $GF(q)$ source symbols, $s_{i-1}=a, s_{i}=b$ and
$s_{i+1}=c$. The prior $Pr(s_{i}=b)$ can be achieved by the
formula\footnote{This prior updating equation is slightly
different then the one originally suggested by Kfir and
Kanter~(\cite{BibDB:KfirEtAl}, Eq. (11)). We find this equation
empirically better.} \BE\label{eq_prior1}
P_{i}^{b}=\sum_{a,c=0}^{q-1}Q_{i-1}^{a}\cdot M_{ab}\cdot
M_{bc}\cdot Q_{i+1}^{c}, \EE where $\mM$ is the measured Markov
transition matrix. The complexity of this updating rule is
$\mathcal{O}(q^2)$. Reducing it to $\mathcal{O}(q)$, the
Eq.~(\ref{eq_prior1}) can be rewritten as
 \BE
P_{i}^{b}=\sum_{a=0}^{q-1}Q_{i-1}^{a}\cdot M_{ab}\sum_{c=0}^{q-1}
M_{bc}\cdot Q_{i+1}^{c}. \label{prior2}
 \EE

As for the noise bit ($j>N$), the coefficient is initialized, for
instance in the BSC case, to be
$Q_{j}^{a}=f^{L}(1-f)^{\log_{2}{q}-L}$, where $L$ is the number of
$1$'s in the bits presentation of the symbol $a$. Then we set
$q_{ij}^{a}=Q_{j}^{a}$ for all non-zero elements in the $j$'th
column.

Note that the complexity per single iteration is almost the same
for both updating methods. Hence, the gain in iterations actually
represents the gain in decoding complexity.

\section{Results}
In order to evaluate the nature of BP convergence, our
experimental study consists of decoding various LDPC codes over
$GF(q)$ for both correlated and uncorrelated information sources.
We perform simulations of decoding over the popular BSC, BEC and
the BI-AWGN channel, using various noise levels (\ie, flip rates,
erasure rates and signal to noise ratios, repectively). The noise
levels are chosen to be close to the noise threshold of the code
in order to get substantial convergence time. The KS code
construction is used with $N=10000/\log_{2}{q}$ source symbols.

Table~\ref{table} compares the average convergence and error rates
for the SUS and PUS. A LDPC code of rate $1/3$ is used, although
similar results are obtained also for other code rates, and the
statistics is collected over at least 6000 different samples. The
correlated sources are modeled by adopting typical 2-state (2S)
and 4-state (4S) Markov processes with entropy
1/2.\footnote{Generated by the Markov transition matrices $\left(
\begin{array}{cc}
0.89 & 0.11 \\
0.11 & 0.89 \\
\end{array} \right)
$ and $\left(
\begin{array}{cccc}
0.808022 & 0.0883281 & 0.0130689 & 0.0905813 \\
0.128676 & 0.0514706 & 0.0772059 & 0.742647 \\
0.755814 & 0.108527 & 0.0116279 & 0.124031 \\
0.866667 & 0.0151111 & 0.091556 & 0.0266659 \\
\end{array} \right)$, respectively.} A convergence speed-up factor of $1/2$, in favor
of SUS, appears. This factor is consistent regardless of the $GF$
order and the source correlation. The standard deviation is
relatively small for all cases. Note that this speed-up does not
deteriorate the code's error performance. As our statistics is
collected over $\sim6000$ samples of block size $10^{4}$, we do
not report the exact value for bit error rate (BER) $P_{b}\leq
10^{-6}$.

\begin{figure}

\begin{center}
\rotatebox{270}{\scalebox{0.675}[0.675]{\includegraphics[width=1\textwidth]{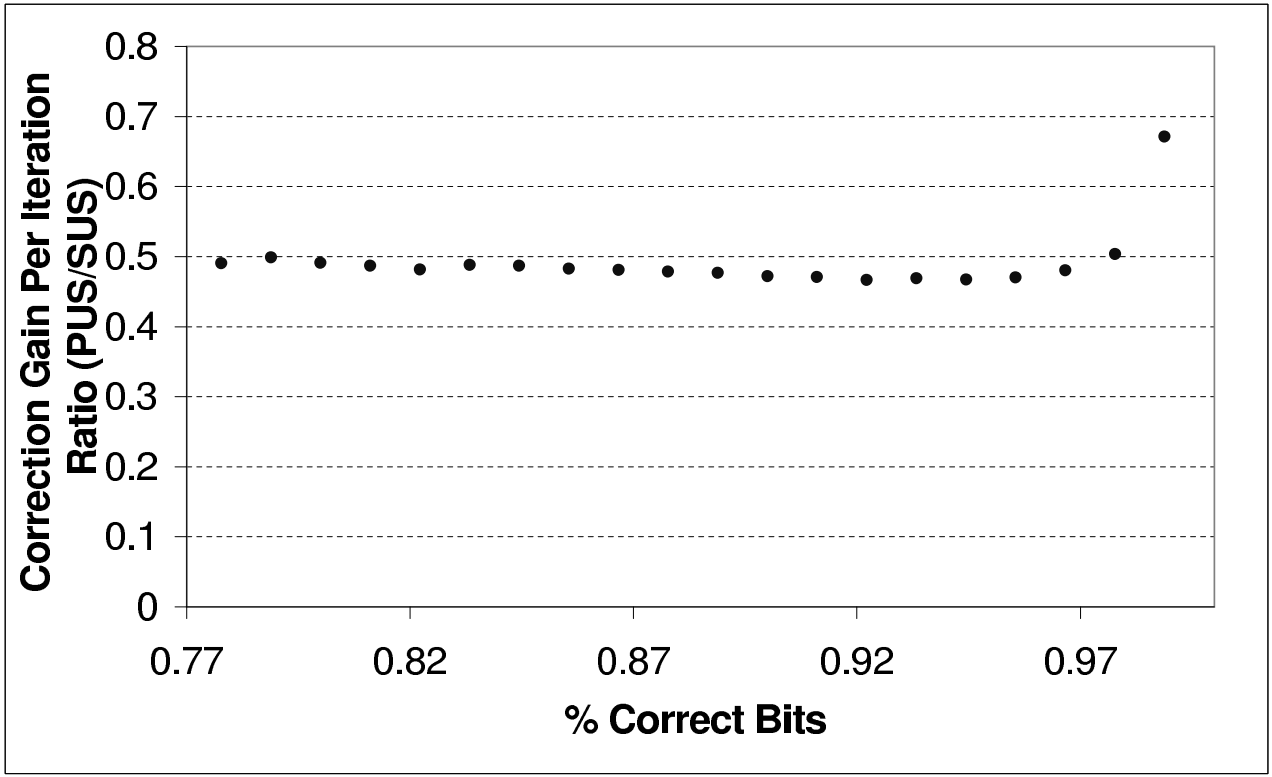}}}

\end{center}
\caption{\label{fig:pim-sim} The relative improvement in correct
bits for PUS/SUS as a function of the current percentage of
correct bits. Using BSC, $N=3333$ (9999 bits), $R=1/3$, $f=0.23$,
$GF(8)$, 10000 samples and a 2-S Markov source.}
\end{figure}

Fig.~\ref{fig:pim-sim} presents the ratio between PUS and SUS in
the percentage of corrected bits per iteration as a function of
the total percentage of correct bits. To be more precise, the
correction gain, \ie the difference between the percentage (w.r.t.
the block's size) of bits corrected in the following iteration and
in the current iteration, $\Delta{P}$, is calculated and the
average ratio for SUS and PUS \BE
\frac{<\Delta{P_{PUS}>}}{<\Delta{P_{SUS}}>}, \EE is drawn as a
function of the percentage of the current correct bits $P$.

It can be easily seen that the one half ratio is preserved through
all the process of decoding, regardless of the decoding dynamics
and the system's state. These results are obtained for
transmitting a 2S-Markov source via a BSC ($f=0.23$) using a
$GF(8)$ rate $1/3$ LDPC code, although similar results are found
for the  BEC and BI-AWGN channel cases, using all other
investigated transmission settings, as listed in
Table~\ref{table}.Notice that the last point in
Fig.~\ref{fig:pim-sim}, indicating the end of the decoding
process, is higher (around $0.7$) since at this final stage the
error correction improvement by SUS can not be twice that achieved
for PUS.

\newpage
\section*{Acknowledgment}
The research of I.K. is supported in part by the Israel Science
Foundation.


\newpage
\begin{landscape}
\begin{table*}
\renewcommand{\arraystretch}{1.3}
\caption{SUS vs. PUS: convergence and error rates} \label{table}
\centering
\begin{tabular}{|c|c|c|c||c|c|c|c|c|c|c|c}
\hline \bfseries Channel & \bfseries GF & \bfseries Source &
\bfseries Noise  & $\begin{array}{c}
\mathbf{<t_{SUS}>}\\(iterations)\end{array}$ & $\begin{array}{c}
\mathbf{<t_{PUS}>}\\(iterations)\end{array}$
 & $\mathbf{\frac{<t_{SUS}>}{<t_{PUS}>}}$
 & $\mathbf{<\frac{t_{SUS}}{t_{PUS}}>}$
 & \bfseries STDEV($\mathbf{\frac{t_{SUS}}{t_{PUS}}}$) & $\begin{array}{c}
\mathbf{BER}(\times 10^{-6})\\SUS,PUS\end{array}$\\
\hline & $2$ & I.I.D. & $f=0.155$ & $25.06$ & $50.73$ & $0.494$ & $0.541$ & $0.080$ & $5.28$ , $5.50$\\
BSC & $4$ & Markov 4S & $f=0.227$ & $28.22$ & $55.66$ & $0.507$ & $0.513$ & $0.055$ & $1.54$ , $1.53$\\
& $8$ & Markov 2S & $f=0.230$ & $25.57$ & $50.04$ & $0.511$ & $0.518$ & $0.073$ & $1.74$ , $1.74$\\
\hline & $2$ & I.I.D. & $\epsilon=0.58$ & $17.60$ & $35.07$ & $0.502$ & $0.503$ & $0.051$ & $<1$ , $<1$\\
BEC & $4$ & Markov 4S & $\epsilon=0.78$ & $22.98$ & $44.54$ & $0.516$ &  $0.527$ & $0.070$ & $1.9$ , $2.0$\\
& $8$ & Markov 2S & $\epsilon=0.76$ & $20.02$ & $40.70$ & $0.492$ & $0.498$ & $0.063$ & $<1$ , $<1$\\
\hline & $2$ & I.I.D. & $\sigma=0.98$ & $20.77$ & $42.64$ & $0.487$ & $0.497$ & $0.026$ & $<1$ , $<1$\\
BI-AWGN & $4$ & Markov 4S & $\sigma=1.64$ & $19.27$ & $37.28$ & $0.517$ & $0.534$ & $0.087$ & $6.2$ , $6.0$\\
& $8$ & Markov 2S & $\sigma=1.66$ & $16.55$ & $32.14$ & $0.515$ & $0.511$ & $0.058$ & $2.6$ , $2.5$\\
\hline
\end{tabular}
\end{table*}
\end{landscape}

\newpage

\end{document}